\documentclass[fleqn,12pt]{wlscirep}
\usepackage[utf8]{inputenc}
\usepackage[T1]{fontenc}
\usepackage{placeins}
\usepackage[normalem]{ulem}
\usepackage{subcaption}
\usepackage{amsmath}
\usepackage{amssymb}
\usepackage{amsfonts}
\usepackage{mathtools}
\usepackage{graphicx}
\usepackage{soul}

\title{Effect of radiation-reaction on charged particle dynamics in a focused electromagnetic wave}

\author[1,2]{Shivam Kumar Mishra*}
\author[1,2]{Sarveshwar Sharma}
\author[1,2]{Sudip Sengupta}
\affil[1]{Institute For Plasma Research, Gandhinagar, Gujarat, India - 382428.}
\affil[2]{Homi Bhabha National Institute, Training School Complex, Mumbai 400094, India}
\affil[*]{mishrasshivam@gmail.com,shivam.mishra@ipr.res.in}

\begin{abstract}
Effect of radiation-reaction force on the dynamics of a charged particle in an intense focused light wave is investigated using the physically appealing Hartemann-Luhmann equation of motion. It is found that, irrespective of the choice of initial conditions, radiation reaction force causes the charge particle to cross the focal region, thereby enhancing the forward energy  
gained by the particle from the intense light wave.
This result is in sharp contrast to the well known result, derived in the absence of radiation reaction forces, where for certain initial conditions the particle reflects from the high intensity region of the focused light wave, thereby losing forward energy.  
These results, which are of relevance to the present day direct laser acceleration schemes of charge particle, also agrees with that obtained using the well known Landau-Lifshitz equation of motion. 
\end{abstract}
\begin{document}

\flushbottom
\maketitle
\thispagestyle{empty}

\section{Introduction}
The relativistic motion of a charged particle placed in an electromagnetic wave is a problem of fundamental interest and is of relevance for investigating the interaction of intense radiation with  matter. The problem was first solved analytically by Landau-Lifshitz \cite{Landau1}  for the case of a charged particle interacting with a relativistically intense, plane monochromatic electromagnetic wave,
using the Hamilton-Jacobi method. The charged particle motion becomes relativistic when the cyclotron frequency  of the particle in the magnetic field of the light wave  ( $\omega_B = e B / m c$; where $e$, $m$, $c$ and $B$ and are  respectively the charge and mass of the charged particle, speed of light and magnetic field of the electromagnetic wave ) becomes of the order (or exceeds) the frequency ($\omega$) of the wave itself {\it i.e.} $\omega_B \geq \omega$. One striking result of this study was, that a charged particle does not gain any energy from a transverse electromagnetic wave\cite{Gibbon, Shebalin,Vikram_sagar_1}. This may be understood from the underlying mechanism of charged particle acceleration in a transverse electromagnetic wave, which is the following. The particle which is initially at rest, first acquires a transverse velocity in the direction of the electric field vector of the light wave. If  $ e E / m \omega c \geq 1$, which is equivalent to $\omega_B \geq \omega$, the magnitude of the transverse velocity reaches close to the speed of light in a time which is a fraction of the wave period. At this moment the Lorentz force on the charged particle due to the magnetic field of the light wave becomes comparable to the force due to its electric field {\it i.e.} ( $e (\vec{v}_\perp/c) \times \vec{B} \sim e \vec{E}$ ), as a result of which the particle starts accelerating thus acquiring a relativistic velocity  along the direction of propagation of the wave. This longitudinal velocity results in a large Doppler shift in the frequency seen by the particle.
As a result the particle remains approximately phase locked with the wave and keeps getting accelerated till the phase of the wave, which is travelling faster than the particle, changes such that the direction of electric field reverses. The particle now starts decelerating and finally comes to a halt. This process then repeats again; as a result the charged particle does not gain any energy from the wave.   \\

This negative result on energy gain can also 
%
%
be understood from the Lagrangian, $L = -m c^2 \sqrt{1 - v^2/c^2} + e \vec{A}.\vec{v} / c$, which represents the interaction of a charged particle with an electromagnetic wave 
propagating along the $z$-direction. 
Here $\vec{A}(t - z/c)$, which is a function of $(t - z/c)$ is the vector potential of the wave and is directed in the transverse direction. Since, the transverse coordinates are cyclic, perpendicular component of canonical momentum $\vec{P_{\perp}}$ is conserved. Also using 
$d H /dt = - \partial L / \partial t = c \partial L / \partial z $, 
we get 
$d H / d t = c (d p_z / dt) $, where $H = \gamma m c^2$ is the Hamiltonian and $p_z$ is the $z$ component of particle momentum  ( The last equality is obtained by using the Lagrange's equation of motion ). This implies that 
$\gamma - p_z/mc = \Delta$ is a constant of motion. Using $\gamma = \sqrt{1 + p_{\perp}^2/m^2c^2 + p_z^2/m^2c^2}$, $p_z$ may be represented in terms of the constants of motion as $p_z/mc = [ 1 - \Delta^2  + (\vec{P_{\perp}} - e\vec{A}/c)^2/m^2c^2 ] / 2 \Delta$. The negative result on energy gain is intimately tied to these two constants of motion, {\it viz.} the perpendicular component of canonical momentum $\vec{P_{\perp}}$ and the parameter $\Delta$. This may be seen as follows: Consider a particle interacting with an incoming electromagnetic wave pulse. Before the pulse reaches the particle position, the particle is at rest implying  $\gamma = 1$ and $\vec{A} = 0$. Therefore the constants of motion are 
$\Delta = 1$ and $\vec{P_{\perp}} = 0$. After the interaction 
$\vec{A} = 0$, and the final energy which is given by $\gamma = 1 + p_{z}/mc$, again turns out to be unity ( as $p_z = 0$ ). Thus the particle does not pick up any energy from the wave. \\

The above discussion suggests that in order for the particle to gain energy from an electromagnetic wave, either one or both the constants of motion must be violated. It is known that a charged particle interacting with an electromagnetic wave in the presence of an external static axial magnetic field, shows unbounded gain in energy when a certain resonance condition is satisfied, {\it i.e.} when the cyclotron frequency of the particle in the external magnetic field matches the Doppler shifted frequency of the wave seen by the particle. This is the well known auto-resonant particle acceleration scheme \cite{kolomenskii1962autoresonance,kolomenskii1963self,sagar2012exact}, where one of the constants of motion {\it viz.} the perpendicular component of canonical momentum is not conserved. Another alternative way of extracting energy from a transverse electromagnetic wave is to focus the wave. This problem of charged particle motion in a focused light wave was first investigated by Feldman et. al.\cite{Feldman} and later in extensive detail by Kaw and Kulsrud\cite{Kaw_Kulsrud}, who introduced a spatial inhomogeneity in the vector potential representing the laser field to model the effect of focusing. The dynamical equations showed that in the focused case, $\vec{P_{\perp}}$ remains constant whereas $\Delta = \gamma - p_{z}$ is no longer a constant of motion, its evolution being dependent on the particle position through the spatial variation of the amplitude of the vector potential. For slow spatial variation of the vector potential, using adiabatic approximation, an approximate expresion for $\Delta$ showing its dependence on particle position was evaluated ( for calculations of $\Delta$ with higher orders of approximation, see ref. \cite{sagar_sengupta_kaw_2013}  ). It was found that the particle gains energy ( along with forward momentum ), when it is kept in the defocused region of the light wave. This region is associated with a monotonic decrease in the parameter $\Delta$. Finally in a recent work \cite{sagar2014effect}, auto-resonant scheme of charged particle acceleration has been studied in the presence of a focused light wave. The charged particle which is initially non-resonant, is accelerated by the focused pulse and brought into resonance; which is then accelerated to very high energies. It was shown that significant amount of energy gain can be achieved at one order of magnitude lesser values of static axial magnetic field and laser intensity. In this scheme both the constants of motion {\it viz.} the perpendicular component of canonical momentum $\vec{P_{\perp}}$ and the parameter $\Delta$ are violated.\\

Recent advances in laser facilities have resulted in renewed interest in  high energy particle acceleration schemes \cite{dla1,dla2}using optical beams produced by Chirped Pulse Amplification method\cite{Strickland,Perry917}. Presently laser light can be focussed to intensities of the order of $10^{23}\, W / cm^2$\cite{shi} ( corresponding electric field $\sim 10^{12} \, V / cm$ ), and in near future intensities are expected to increase by two orders of magnitude or more\cite{shi} ( the corresponding electric field can reach of the order of $\sim 10^{13}\, V / cm$ ). In the context of particle dynamics in such high fields, it can easily be seen that the power radiated  by an electron (or positron) becomes comparable to the instantaneous rate of change of energy of the particle; which in turn implies, that in this scenario, the radiation reaction force becomes comparable to the Lorentz force acting on the particle\cite{Shen,Shen1,hadad}. Earlier studies on laser driven charged particle acceleration as discussed in the previous paragraph had neglected radiation reaction effects. It turns out that in the presence of radiation reaction effects both the constants of motion {\it i.e.} $\vec{P}_{\perp}$ and $\Delta$ are violated\cite{hadad,mishra_epjst}. Recent studies\cite{Piazza-prl,Piazza1,hadad,mishra_epjst,gong2019radiation} have shown that, although counter-intuitive, radiation reaction forces actually lead to energy gain by the particle. 
Thus, at high laser intensities, the study of charged particle dynamics in a focused laser field with the inclusion of radiation reaction effects is important from the point of view of direct laser acceleration schemes \cite{dla1,dla2}. Previous studies in this area had either neglected radiation reaction effects\cite{Kaw_Kulsrud,Feldman} or focusing effects\cite{hadad,Piazza-prl,Piazza1,sagar22,mishra_epjst}. The present article is devoted to the study of charged particle dynamics  under the combined effect of focusing and radiation reaction forces. \\

The fundamental equation which, within the framework of classical electrodynamics,  self-consistently takes into account 
the effect of radiation reaction forces is the covariant Lorentz-Abraham-Dirac (LAD) equation of motion\cite{Lorentz, Dirac1,Dirac2, Abraham}.
LAD equation takes into account both the radiated energy ( electromagnetic fields which are irreversibly transported to infinity ) and the Schott energy which is the field energy localized at the particle and can be exchanged with its mechanical energy. 
However, it is well known that the LAD equation suffers from unphysical problems like pre-acceleration and runaway solutions. These unphysical problems have been widely discussed in literature\cite{Griffiths,jackson,roharlic1,Landau1,rohrlich3,Kasher,bhabha}. Apart from the LAD equation of motion, there exists other equations of motion {\it viz.} F. O'Connell(FOC)\cite{FOC1,FOC2,FOC3}, Eliezer\cite{Eliezer}, Landau-Lifshitz\cite{Landau1}, 
M. O. Papas\cite{MO}, Caldirola\cite{Caldirola}, Hartemann Luhmann\cite{Hartemann}, Yaremko\cite{Yaremko}, Sokolov \cite{Sokolov:2009qku,Sokolov,SokolovPRE} et. al.,  which takes into account the radiation reaction effects and have been derived using different approaches. In our work, we have used the following well known equations for studying the effect of radiation radiation on a charged particle interacting with a focused laser pulse.
One of them is the widely used Landau-Lifshitz equation of motion, which is perturbatively derived from the LAD equation, by substituting the time derivatives of four velocity in the radiation reaction term, by the Lorentz force term. Landau-Lifshitz equation is extensively used in literature, for studying the effect of radiation reaction on accelerated charged particles \cite{hadad,Piazza-prl,sagar22}. Although derived using a perturbative method, the Landau-Lifshitz equation does not suffer from the above mentioned unphysical problems. Besides the Landau-Lifshitz equation, there exists another physically appealing model which also does not suffer from unphysical problems; it is the Hartemann-Luhmann equation \cite{Hartemann}. The derivation of Hartemann-Luhmann equation is based on the simple fact that the angular distribution of radiation emitted by an accelerating charged particle, becomes highly asymmetric due to relativistic Doppler effect. As a result, radiation is mainly emitted in the direction of the velocity vector of the particle resulting in a net reaction force in a direction opposite to the velocity of the  particle. By calculating the radiation pressure force on a accelerating charged sphere of radius ``$R$'' and then taking the limit $ R \rightarrow 0$, final expression of radiation reaction force acting on a accelerating charged particle was obtained. An outcome of this approach is that the radiation reaction term contains the effects due to far field alone ( electromagnetic fields which are irreversibly radiated away )\cite{Hartemann,Bellotti}. The Hartemann-Luhmann equation, thus can also be derived directly from the LAD equation, by neglecting the term due to Schott energy in comparison to the term due to radiated energy in the expression for radiation reaction\cite{mishra_epjst}.\\

In the present work, we use the Hartemann-Luhmann equation for studying the dynamics of a charged particle in an intense focused light wave. This equation is numerically solved using a MATLAB based code, used for solving implicit differential equations. It is found that, independent of the choice of  field configuration, with the inclusion of radiation reaction term in the equation of motion results in a monotonic decrease in the parameter $\Delta$, in the radiation reaction dominated regime.
This, as mentioned above ( and also shown explicitly in our earlier work\cite{mishra_epjst} ), is associated with energy gained by the particle from the electromagnetic wave, along with gain in forward longitudinal momentum. It is found that, irrespective of the choice of initial conditions, radiation reaction force causes the paricle to cross the focal region, thereby enhancing the forward energy gained by the particle from the intense light wave. This result is in sharp contrast to the well known result by Kaw et. al.\cite{Kaw_Kulsrud}, where for certain initial conditions, the particle reflects back from the high intensity regime of the focused light wave, thereby losing forward energy. The schematic diagram shown in Fig. (1), summarizes the above mentioned observations.
 \begin{figure}
 	\centering
 	{\includegraphics[width = 7 in]{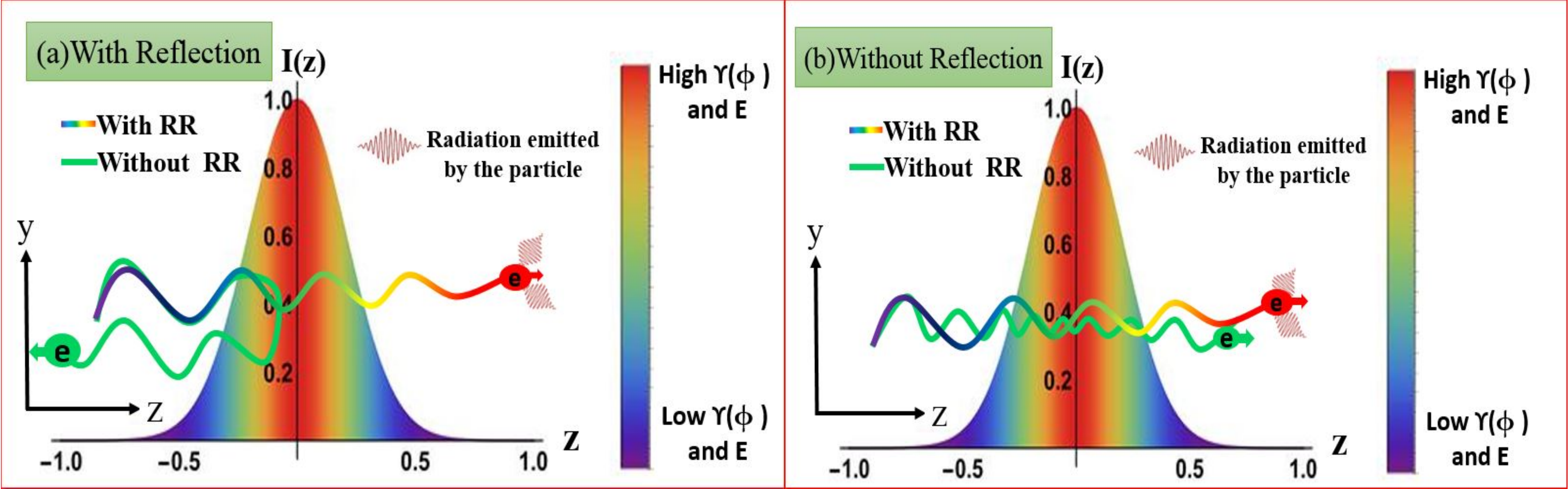}}
 	\caption{This schematic diagram compares the charged particle dynamics in a focused light wave in the absence / presence of radiation reaction ( RR ) effects. The variation of laser intensity ( hence laser electric field ) along the direction of propagation ( z - axis ) due to focusing is shown by a colour bar. Sub-figures (a) and (b) represent particle trajectories with initial conditions which respectively lead to reflection and transmission  through the focal point  in the absence of radiation reaction forces ( the green wavy lines in sub-figures (a) and (b) ). It is shown that inclusion of radiation reaction forces results in transmission through the focal point in both cases, irrespective of the choice of initial conditions. On transmission, the increase in energy of the particle is also shown by change in colour. 
}
 	\label{3}
 \end{figure}
\\

The organization of the paper is as follows: In section II of this article,  we present the basic governing equations describing the dynamics of a charged particle in a focused light wave, including radiation reaction effects. Beginning from the LAD equation, we point out the various approximations which lead to the Landau-Lifshitz and the Hartemann-Luhmann equation of motion. Hartemann-Luhmann equation is then applied to the case of a focused electromagnetic wave.
For the sake of completeness, 
and to get an insight into the charge particle dynamics in a focused 
light wave, we then discuss the reflection condition obtain by Kaw et. al.\cite{Kaw_Kulsrud}, without taking account of radiation reaction effects.
Finally in this section, using the Hartemann-Luhmann equation and based on the equation for the parameter $\Delta$, we indicate the implications which occur due to the inclusion of radiation reaction effects.
Numerical results, obtained by solving the Hartemann-Luhmann equation, describing the dynamics of a charged particle interacting with a focused wave train is presented in sections III. 
All the results obtained by solving the Hartemann-Luhmann equation of motion has been further compared with the results obtained by solving the Landau-Lifshitz equation of motion. The obtained results show that the energy gain by the particle is independent of the model equation used. Finally in section V, we present a summary of our results.

\section{Governing Equations} \label{gov_eqns}
The most fundamental equation which describes the interaction of a spinless point charged particle with an electromagnetic wave, including radiation reaction effects is the Lorentz-Abraham-Dirac ( LAD ) equation, which may be written in dimensionless covariant form as
\begin{equation}\label{LAD}
\dot{u}^{\alpha} = F^{\alpha \beta} u_{\beta} + \tau_0 \left( \delta^{\alpha}_{\beta} - u^{\alpha}u_{\beta} \right)\ddot{u}^{\beta}
\end{equation}
where $F^{{\alpha}{\beta}} \rightarrow e F^{{\alpha}{\beta}}/m\omega c$ is the electromagnetic field tensor, $u^{\alpha} \rightarrow u^{\alpha}/c = \gamma (1, \vec{\beta}) $ is the four-velocity, $\gamma = \sqrt{1 - v^2/c^2}$,  $\tau_{0} \rightarrow \omega \tau_0 =( 2 / 3)(e^2 k / m c^2)$
( $\omega, \, k$ being the frequency and wave number of the electromagnetic wave ) and the dot represents differentiation with respect to proper time $\tau$. We use the metric tensor $(+1,-1,-1,-1)$. The two terms in the expression for radiation reaction respectively represent the radiation reaction force due to Schott energy and the radiated energy. It is the presence of Schott term in the LAD equation which leads to unphysical problems like pre-acceleration and runaway solutions. Following Landau-Lifshitz, if the  dimensionless field amplitude $a_0$ satisfies the inequality $\omega \tau_0 a_0 \ll 1$ in the instantaneous rest frame of the charged particle, then the acceleration term in the radiation reaction can be replaced by the Lorentz force term; which leads to the Landau-Lifshitz equation of motion as 
\begin{equation}\label{LL}
\dot{u}^{\alpha} = F^{\alpha \beta} u_{\beta} + \tau_0 \left( \delta^{\alpha}_{\beta} - u^{\alpha}u_{\beta} \right) \left( F^{\alpha \beta}_{, \delta}u^{\delta}u_{\gamma} + F^{\beta \gamma}F_{\gamma \delta} u^{\delta} \right) 
\end{equation}
where 
$F^{\beta \gamma}_{, \delta}u^{\delta} = (d/d\tau) F^{\beta \gamma}$. Approximation of the Schott term, thus eliminates the unphysical problems associated with the LAD equation. Besides the above approximation,  it can also be seen from the LAD equations that the Schott term is of order unity, whereas the radiated term is of order $\sim \gamma^2$. So in the ultra-relativistic case, where radiation reaction becomes important, Schott term may be neglected in comparison with the radiated term, thus leading to the Hartemann-Luhmann equation\cite{Hartemann} as ( For a detailed discussion on this approximation see ref.\cite{mishra_epjst} ). 
\begin{equation}\label{HL}
\dot{u}^{\alpha} = F^{\alpha \beta} u_{\beta} + \tau_{0} \dot{u}^{\beta}\dot{u}_{\beta} u^{\alpha} 
\end{equation}
Below, for the case of a charged particle interacting with a focused electromagnetic wave, we work with the Hartemann-Luhmann equation.  \\

Consider the motion of a charged particle in a focused electromagnetic wave. The effect of focusing is modelled by taking the amplitude of the vector potential to be a function of space, as $\vec{A} = a(z)\vec{\zeta}(\phi)$, where $a(z)$ is the spatially dependent amplitude of the vector potential, z being the direction of propagation, and 
$\vec{\zeta}(\phi) = \theta(\phi) P(\phi)$ which depends only on the phase $ \phi=\omega t - k z $, is a product of an oscillatory part $P(\phi)$ and a pulse shaping envelope $\theta(\phi)$ ( for a wave train $\theta(\phi)$ is unity ). Transverse nature of the electromagnetic wave implies that the vector potential $\vec{A}$ ({\it i.e.}
$\vec{\zeta}(\phi)$) lies in the x-y plane ( plane perpendicular to the direction of propagation ). Here $a(z)$ is chosen in such a way that intensity of the laser light is maximum at the focal point and decreases on either side of it ( as shown in Fig. (1) ). The corresponding electric and magnetic field of the electromagnetic wave may be written as
$\vec{E} = -( 1 / c ){\partial \vec{A}}/{\partial t} = -(\omega/c)a(z)\vec{\zeta^{'}}(\phi) $ and 
$\vec{B}= \vec {\nabla } \times \vec{A} = - a(z)\{\vec{k} \times \vec{\zeta^{'}}(\phi)\} + \vec{\nabla} a(z) \times \vec{\zeta}(\phi)$ respectively, where prime represents derivative with respect to $\phi$ and $\vec{k} = k \hat{z}$, $\omega$ are respectively the wave vector and frequency of the electromagnetic wave. Now the temporal and spatial components of the Hartemann-Luhmann equation of motion\cite{Hartemann, mishra_epjst} in dimensionless form can be written as
\begin{equation}\label{energy_hl} 
\frac{d \gamma}{dt} = \vec{\beta}.\vec{E} - \tau_{0} \gamma^{4} \left\{\left(\frac{d\vec{\beta}}{dt}\right)^{2} + \gamma^{2} \left(\vec{\beta}.\frac{d\vec{\beta}}{dt}\right)^{2}\right\}
\end{equation}
\begin{equation}\label{momentum_hl}
\frac{d \vec{p}}{dt} = \left(\vec{E} + {\vec{\beta}} \times \vec{B}\right) - \tau_{0} \gamma^{4} \left\{ \left(\frac{d\vec{\beta}}{dt}\right)^{2} + \gamma^{2} \left(\vec{\beta}.\frac{d\vec{\beta}}{dt}\right)^{2}\right\} \vec{\beta}
\end{equation}
where the normalization used is $t \rightarrow \omega t$, $\vec{r}\rightarrow k \vec{r}$, 
$\vec{p} \rightarrow {\vec{p}}/{mc}$, $\vec{\beta} = \vec{v}/c$, $ \vec{A} \rightarrow {e\vec{A}}/{mc^{2}}$, $ \vec{E} \rightarrow {e \vec{E}}/{m \omega c}$, $\vec{B} \rightarrow {e \vec{B}}/{m \omega c} $, $\tau_0 \rightarrow \omega \tau_0$ and $dt = \gamma d \tau$. Substituting the normalized expressions for $\vec{E}$ and $\vec{B}$ in the above equations, the energy equation, and the parallel and perpendicular components of the equation of motion ( parallel and perpendicular to the direction of propagation of the wave ) may respectively be written as 
\begin{equation}\label{energyeq_rr} 
\frac{d \gamma}{dt} =  -a \vec{\beta}.\vec{\zeta}' - \tau_{0} R_{h}
\end{equation}
\begin{equation}\label{pzeqrr}
\frac{d p_{z}}{dt} = - a \vec{\beta}.\vec{\zeta}' + \frac{da}{dz} \vec{\beta}.\vec{\zeta} - \tau_{0} R_{h} \beta_{z}
\end{equation}
\begin{equation}\label{pperpeqrr}
\frac{d \vec{p}_{\perp}}{dt} = -a \left( 1 - \beta_z \right) \vec{\zeta}'  - \frac{da}{dz}\beta_z\vec{\zeta} - \tau_{0} R_{h} \vec{\beta}_{\perp}
\end{equation}
where $\beta_{z}$ is the z-component of velocity, $\gamma = \sqrt{1 + p_z^2 + p_\perp^2}$ and $R_h = \gamma^4(\dot{\beta}^2 + \gamma^2 (\vec{\beta}.\dot{\vec{\beta}})^2 )$. 
To get an insight into the dynamics of the charged particle, we first consider the case without the radiation reaction term 
{\it i.e. $\tau_0 \rightarrow 0$}. This was the case first considered by Kaw et. al. \cite{Kaw_Kulsrud}, which we present here for the sake of completeness.  Using $(1 - \beta_z)\vec{\zeta}^{'} = d\vec{\zeta}/dt$ and $(da/dz)\beta_{z} = da/dt$, Eq. (\ref{pperpeqrr}) may be integrated, which leads to conservation of perpendicular component of canonical momentum as
\begin{equation}\label{eqn6}
\vec{p}_{\perp} + a\vec{\zeta} = \vec{P}_{\perp 0}
\end{equation}
where $\vec{P}_{\perp 0}$ is a constant of motion. It is clear from the above equation that, in order to conserve $\vec{P}_{\perp 0}$, the amplitude of perpendicular component of particle momentum ( $\vec{p}_{\perp}$ ) increases as the particle moves towards the focal point ( regions of increasing $a$ ) and decreases as it moves away from the focal point ( region of decreasing $a$ ). Further defining $\Delta = \gamma - p_z$, the longitudinal component of momentum may be written in terms of $\vec{P}_{\perp 0}$ and $\Delta$ as 
\begin{equation}\label{pz}
p_z = \frac{1 - \Delta^2}{2 \Delta} + \frac{(\vec{P}_{\perp 0} -  a\vec{\zeta})^2}{2 \Delta}
\end{equation}
Thus knowledge of initial conditions ( {\it i.e.} $\vec{P}_{\perp 0}$  and $\Delta$ ), completely specifies the perpendicular and longitudinal component of momentum of the particle. The equation for $\Delta$ may be obtained by subtracting Eq. (\ref{pzeqrr}) from Eq. (\ref{energyeq_rr}) as
\begin{equation}\label{deltaeq}
\frac{d\Delta}{dt}= - \frac{da}{dz} \vec{\beta}.\vec{\zeta} - \tau_{0} R_{h} \left( 1-\beta_z \right)
\end{equation}
It is clear from above that, in the absence of radiation reaction 
($\tau_0 \rightarrow 0$) and in the unfocused case ({\it i.e.} $da/dz = 0$), $\Delta$ is a constant of motion and hence is entirely determined by the initial conditions. In the present case, $\Delta$ is no longer a constant of motion and varies with the position of the particle. Taking $\tau_0 = 0$ and assuming the amplitude of the vector potential ``$a$'' to be a slowly varying function of position ``$z$'', we may write $\dot{p_z} \approx \overline{\dot{p_z}}$ and 
$\dot{\Delta} \approx \overline{\dot{\Delta}} \approx (d \Delta / dz) \overline{p_z} $, where the overline represents averaging over fast variation and the ``dot'' represents derivative with respect to proper time 
($\tau$; and  $dt = \gamma d \tau$). Thus averaging over fast variation, the equation for slow temporal variation of the longitudinal momentum of the particle ( from Eq. (\ref{pzeqrr}) ), may now be written as
\begin{eqnarray}\label{eqn8}
\frac{d {p_{z}}}{d\tau} &\approx& - \overline{a \vec{p}.\vec{\zeta}'}  + \overline{\frac{da}{dz} \vec{p}.\vec{\zeta}} \nonumber \\
& \approx & -a \overline{\vec{P}_{\perp 0}.\vec{\zeta}'} + 
a^2 \overline{\vec{\zeta}.\vec{\zeta}'}+ \frac{da}{dz}\overline{\vec{P}_{\perp 0}.\vec{\zeta}} - a \frac{da}{dz}\overline{\zeta^{2}} \nonumber\\
& \approx & -\frac{1}{2} a \frac{da}{dz} 
\end{eqnarray}
where $\gamma \vec{\beta} = \vec{p}$ and in the second step, we have used Eq. (\ref{eqn6}) and the fact that $\vec{\zeta}$ is perpendicular to the direction of propagation of the wave ({\it i.e.} z). Taking $\vec{\zeta}(\phi)$ to be a purely sinusoidal function of $\phi$ 
{\it i.e. } $\theta(\phi) = 1$, we get $\overline{\zeta^{2}} = 1/2$ and the other terms vanish. The behaviour of particle dynamics may be understood from the above averaged equation in the following way. Eq. (\ref{eqn8}) shows, that in the region before the focal point $da/dz$ being positive, as the particle moves forward ( due to $\vec{v} \times \vec{B} $ force ) the average value of $p_z$ decreases monotonically with time, and for certain initial conditions may even vanish before reaching the focal point. For such cases, the particle will reflect from the high intensity region, and thus it will not gain any forward energy from the light wave.  On the other hand, if the particle either starts from the defocused region or from the focused region with sufficient amount of initial forward momentum so that it crosses the focal point and reaches the defocused region, it will gain a large amount of forward momentum ( hence energy ) from the defocused region. This is because, in the defocused region, $da/dz$ being negative, $p_z$ increases monotonically with time. This indicates that in order to extract substantial amount of forward energy, the particle must start from the defocused region. The above qualitative discussion on particle dynamics may be put on a quantitative footing by integrating the averaged equation for $\Delta$ as follows. Using  $\dot{\Delta} \approx \overline{\dot{\Delta}} \approx (d \Delta / dz) \overline{p_z} $, we get 
\begin{eqnarray}\label{eqn9}
\frac{d\Delta}{d z} &\approx& - \frac{da}{dz} \frac{\overline{\vec{p}.\vec{\zeta}}}{\overline{p_z}} \nonumber \\
& \approx & \frac{1}{2} a \frac{da}{dz} \frac{2 \Delta}{1 - \Delta^2 + P_{\perp 0}^2 + a^2 / 2}
\end{eqnarray}
where, as before, in the second step we have used Eq. (\ref{eqn6}) and 
$\overline{p_z}$ is evaluated using Eq. (\ref{pz}) taking
$\vec{\zeta}(\phi)$ to be a purely sinusoidal function of $\phi$.
Multiplying both sides by $\Delta$, Eq. (\ref{eqn9}), immediately gives
\begin{equation}\label{eqn12}
\frac{d \Delta^2}{d a^2} \approx - \frac{\Delta^2}{\Delta^2 - P_{\perp 0}^{2} - a^{2}/2 - 1}
\end{equation}
which when integrated yields $\Delta$ as a function of position $z$ as, 
%
\begin{equation}\label{Delta}
\Delta  \approx 1 + \frac{P_{\perp 0}^2}{2} + \frac{a(z_0)^2}{4} - \left[ \frac{1}{4} P_{\perp 0}^2 \left( P_{\perp 0}^2 + a(z_0)^2 \right) + \frac{1}{16} a(z_0)^4 + \frac{1}{2} \left( a(z_0)^2 - a^2 \right) \right]^{1/2}
\end{equation}
where $a(z_0)$ is the value of $a$ at the initial position $z_0$ from where the particle is assumed to start from rest ( {\it i.e.} 
$\Delta = 1$ ). From the above expression of $\Delta$ it is clear that, as the particle moves towards the focal point ( {\it i.e.} increasing a ), the term in the square root decreases ($\Delta$ increases) and will vanish at a particular position ($z_{ref}$) before the particle reaches the focal point, provided the following inequality is satisfied 
\begin{equation}\label{con1}
a_{z_{ref}}^{2} = \left[\frac{1}{2} P_{\perp 0}^2 \left( P_{\perp 0}^2 + a(z_0)^2 \right) + \frac{1}{8} a(z_0)^4 + a(z_0)^2 \right] < a_0^2 
\end{equation}
where $a_0$ and $a_{z_{ref}}$ are the amplitude of the vector potential at the focal point and at the reflection point ($z_{ref}$) respectively.  Beyond this position, $\Delta$ becomes complex, which physically implies that the particle cannot go into regions of $ a > a_{z_{ref}} $ and will thus reflect from $z_{ref}$. Since $a(z)$ is known, the point of reflection $z_{ref}$ may be evaluated from the value of $a_{z_{ref}}$. As stated before, the average forward longitudinal momentum decreases as the particle moves towards the focal point and vanishes at the point of reflection ($z_{ref}$). This may be seen by substituting the expression for $\Delta$ at the reflection point 
which is $\Delta_{ref} = 1 + P_{\perp 0}^2/2 + a(z_0)^2/4$, in the expression for average longitudinal momentum $\overline{p_z} = ( 1 - \Delta_{ref}^2  + P_{\perp 0}^2 + a_{z_{ref}}^2/2 ) / 2 \Delta_{ref}$, which gives $\overline{p_z} = 0$. On the contrary, if the particle starts with the initial conditions, such that $a_{ref}^2 > a_0^2$, $\Delta$ will remain real during the entire motion and the particle will cross the focal point and enter the defocused region. In the defocused region, with decreasing $a$, $\Delta$ decreases and eventually becomes less than unity for $a < a(z_0)$.  This results in large gain in forward longitudinal momentum as seen from Eq. (\ref{pz}) with concomitant gain in energy. Thus, as also stated before, the particle gains energy from the defocused region of the electromagnetic wave. 

To predict, how the above dynamical behaviour of the particle changes in the presence of radiation reaction forces,
we note that the radiation reaction term ``$\tau_0 R_{h} (1-\beta_{z})$'' in Eq. (\ref{deltaeq}) is always a positive number. Therefore in the radiation reaction dominated regime, where the radiation reaction force dominates over the Lorentz force, Eq. (\ref{deltaeq}) shows that the parameter $\Delta$, monotonically decreases with time. This behaviour of the parameter 
$\Delta$ along with the definition  of
$\gamma = \sqrt{1 + p_{\perp}^{2} + p_z^2}$ implies that for a particle starting from rest, increase in its forward  energy is associated with the increase in its longitudinal momentum. This is in conformity with the intuitive understanding evolved for the case which is without radiation reaction effects. In the next section, we present numerical solutions to Hartemann-Luhmann and Landau-Lifshitz equation of motion using different initial conditions.
\section{Dynamics of a charged particle in a focused light wave} 
\label{dynamics}
In this section, we present numerical solution to Eqns. (\ref{energyeq_rr}) - (\ref{pperpeqrr}), obtained  using an in-house developed MATLAB based test particle code. These equations are solved for a particle interacting with a 
focused light wave. The vector potential corresponding to a focused wave train may be written as,
\begin{align}\label{vec_p}	
A(\phi,z) = a \left( \delta   \cos(\phi) \hat{x} +  g \sqrt{1-\delta^{2}} \sin(\phi) \hat{y} \right)
\end{align}	
where to model the effect of focusing, the functional form of amplitude $a(z)$ is taken as,
\[
	a(z)= 
	\begin{dcases}
		a_0 ( 1 + \epsilon z),& \text{for } z\leq 0\\
		a_0 ( 1 - \epsilon z),  & \text{for } z\geq 0
	\end{dcases}
\]
\begin{figure}
\centering
{\includegraphics[width = 7in]{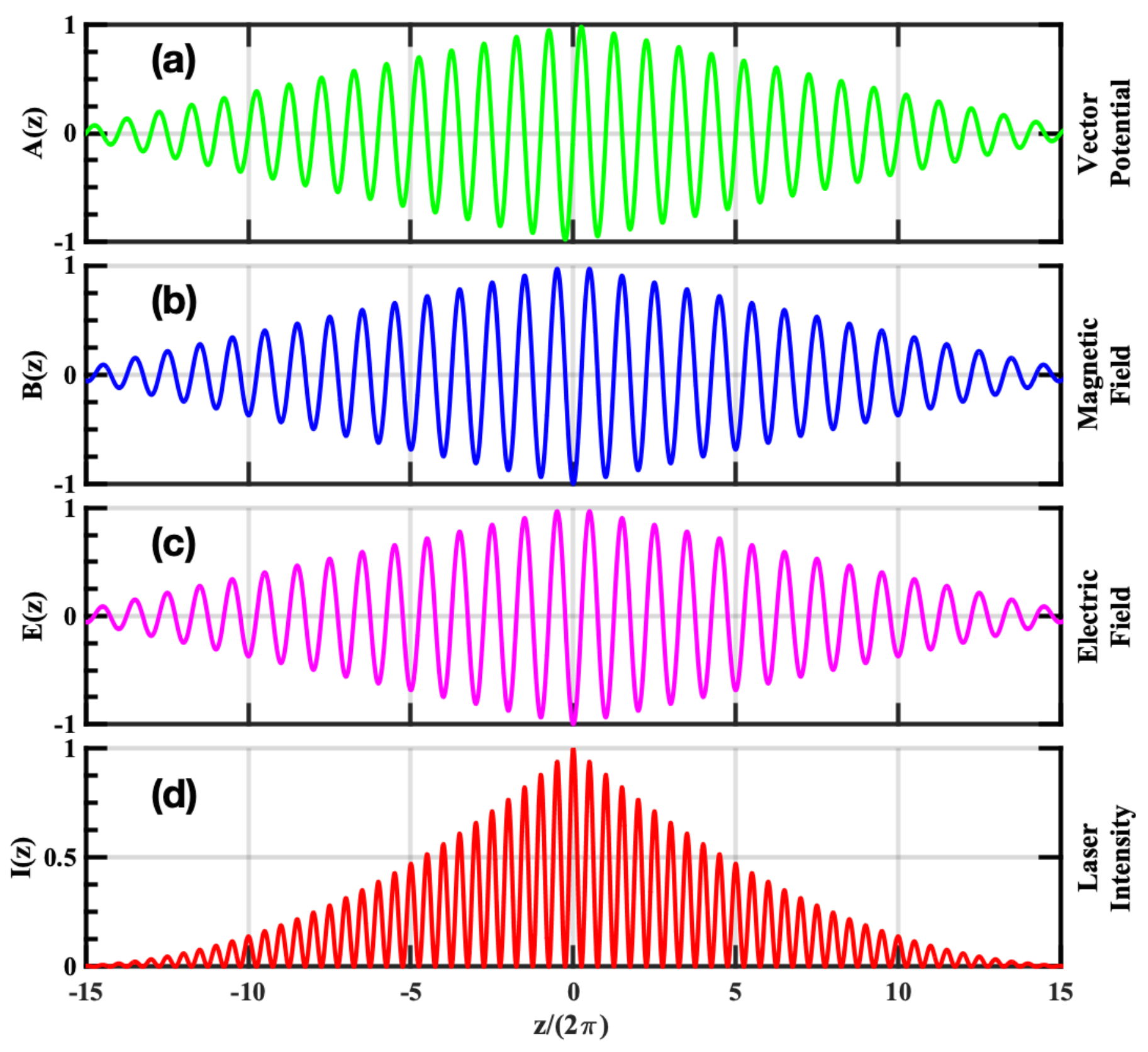}}
\caption{In this schematic diagram the sub-figures (a) - (d), respectively represent the  variation of the vector potential, magnetic field, electric field and intensity of the laser  with respect to the z-coordinate. The figures show that the laser fields and laser intensity are maximum at the focal point ( $z=0$ ) and decreases on either side of it.}
\label{schematic_2}
\end{figure}
Here, $\epsilon \ll 1$ and $\delta \in [0,1]$ where $\delta = 0,\,1$ and $\delta = 1/\sqrt{2}$ correspond to linear and circular polarization respectively; and $g = \pm 1$ respectively correspond to right and left handed polarization. To get a feel for the field and intensity profiles for a focused wave train, normalized laser fields and the laser intensity corresponding to the normalized vector potential given by Eq. (\ref{vec_p}) are plotted in Fig. (2), for $\delta=0$, $a_{0} = 1$ 
and $\epsilon \sim 0.03$. In this schematic diagram, the sub-figures (a), (b), (c) and (d) respectively represent the variation of vector potential, magnetic field, electric field and intensity of the laser over $\sim 30$ cycles of the laser period. The laser field is maximum at the focal point ( $z=0$ ) and decreases on either side of the focal point. \\

We now present results obtained for a charged particle interacting with a linearly polarized ( $\delta = 0$ ), intense, focused wave train ( for our numerical work, we have chosen $\epsilon \sim 10^{-4}$ ). The intensity of the wave at the focal point is chosen to be  $a_{0}=1000$ ( $\sim 10^{24} \, W/cm^2$ ), and $\tau_{0} \approx 1.8 \times 10^{-8}$ ( for a $1 \mu$ wavelength laser ). For our numerical work, we have used two sets of initial conditions. For the first set, the initial conditions are chosen in such a way that the inequality given by Eq. (\ref{con1}) is satisfied. For this set, the particle is assumed to start from rest ( $\vec{p}=0$ ) from the initial position $z_{0}=-9500$. The phase of the wave initially seen by the particle is chosen such that the perpendicular canonical momentum vanishes ( {\it i.e.} $\vec{P}_{\perp 0} = 0$ ) and $a(z_0) = 50$. For these initial conditions $a_{z_{ref}} \sim 885.3 < a_0$; thus as per Eq. (\ref{con1}) it is expected that during the motion, the parameter
$\Delta$, will become complex and the particle will eventually reflect from the high intensity region. For the second set, the choice of initial conditions is such that the above mentioned inequality (Eq. (\ref{con1}) ) is violated. For this set, the particle is assumed to start from rest ( $\vec{p}=0$ ), from the initial position $z_{0}=-9450$, which is now placed closer to the focal point than that in the previous case. As before, the phase of the wave initially seen by the particle is chosen in such a way that the perpendicular canonical momentum vanishes ( {\it i.e.} $\vec{P}_{\perp 0} = 0$ ) and $a(z_0) = 55$. For these initial conditions and for the same laser and focusing parameters, $a_{z_{ref}} \sim 1070.9 > a_0$; thus as per Eq. (\ref{con1}), the parameter $\Delta$ will remain real throughout the motion and the particle is expected to pass through the focal point. \\

For the first set of initial conditions, the numerical results are presented in Figs. (3) and (4). In Fig. (3), the sub-figures (a), (b), 
(c) and (d) respectively represent the longitudinal ( $p_z$ ) and transverse ( $p_{y}$ ) component of particle momentum, the energy and the parameter $\Delta$; and Fig. (4) represents the trajectory of the particle. 
We first discuss the particle dynamics in the absence of radiation reaction effects. In Fig. 3 (a) -  (c) and Fig. (4), the red and green curves, respectively represent the forward and the reflected motion of the particle in the absence of radiation reaction effects ( values correspond to y-axis on the left ). The red curves in Fig. 3 (a) and (b) respectively show that as the particle approaches the focal point ( increasing $a$ ), the average longitudinal momentum decreases monotonically and approaches zero, whereas the amplitude of the transverse momentum simultaneously increases. After reflection ( represented by the green curves ), as the particle moves away from the focal point ( decreasing $a$ ), the absolute value of average longitudinal momentum increases and the particle leaves the focal region with a finite value of longitudinal momentum in the opposite direction, whereas the amplitude of the transverse momentum decreases and eventually goes to zero. These numerical results are respectively in conformity with Eqs. (\ref{eqn8}) and (\ref{eqn6}). The corresponding energy is shown in Fig. 3(c), which shows that for our choice of parameters the average energy which remains with the particle after reflection is around $\gamma \sim 600$ ( $\sim 0.3 \, GeV$ ). It is to be noted that the average energy does not change throughout the motion, which is in conformity with the average of Eq. (\ref{energyeq_rr}) ($\overline{d \gamma  / d \tau} \approx 0$, with $R_h = 0$). Finally, Fig. 3(d) shows the evolution of the parameter $\Delta$ as a function of $z$ coordinate, which continuously increases throughout the motion as shown by the red curve. The green curve on top of the red curve is a plot of the analytical expression of $\Delta$ as given by Eq. (\ref{Delta}), which clearly shows an excellent match with the numerical result. The particle trajectory is presented in Fig. (4) ( red and green curves, respectively represent the forward and the reflected motion of the particle, with values corresponding to the y-axis on the left ), where sub-figure 4(a) represents the trajectory in configuration space and sub-figure 4(b) presents its temporal evolution. The reflection of the particle at $\phi \sim 0.5 \times 10^{4}$ and $z = - 1.1 \times 10^{3}$  is clearly visible in sub-figure 4(b).\\

For the same set of initial conditions, in the presence of radiation reaction effects the particle dynamics which is now governed by the Hartmemann-Luhmann equation, exhibit dramatic changes. ( See the blue curve in sub-figures Fig. 3(a), 3(c), 3(d) and Fig. 4(a) with values corresponding to y-axis on the right ). Initially when the radiation reaction effects are weak, the average longitudinal momentum $\overline{p_z}$ remains almost constant ( see blue curve in Fig. 3(a); in fact it shows a slight decrease, which is in agreement with the behaviour governed by the Lorentz force terms in Eq. (\ref{pzeqrr}) ), and later increases monotonically when the radiation reaction term starts dominating over the Lorentz force term. This dominance of the radiation reaction term over the Lorentz force term can also be seen from the evolution of the parameter $\Delta$ ( see Fig. 3(d) ) which shows that $\Delta$ which was initially increasing, starts monotonically decreasing when the intensity becomes sufficiently large  ( $a \sim 6.3 \times 10^{2}$, intensity $\sim 4 \times 10^{23}\, W/cm^2$ ) which happens when the particle reaches around $z \sim -0.6 \times 10^{4}$. From this location onwards, the radiation reaction term starts dominating over the Lorentz force term. Simultaneously, from  around the same location {\it i.e. $z \sim -0.6 \times 10^{4}$ }, the  average longitudinal momentum begins to increase monotonically and  
the particle eventually passes through the focal point with a  finite amount of longitudinal momentum. The transverse momentum on the other hand shows a behaviour which is similar to the earlier case, {\it i.e.}when radiation reaction effects are absent or weak . The amplitude of the transverse momentum of the particle increases as it approaches the focal point and diminishes as it passes through the focal point, eventually becoming zero as it exits the focal region ( see blue curve in Fig. 3(b) ). Therefore the final energy gain as seen in Fig. 3(c) (blue curve), is entirely due to the net gain in longitudinal momentum. This is in agreement with the understanding developed using Eq. (\ref{deltaeq}), according to which, in the radiation reaction dominated regime, a monotonic decrease in the parameter $\Delta$ implies energy gain along with increase in forward longitudinal momentum. The energy gain is found to be $\gamma \sim 1.4 \times 10^{4}$ ($\sim 7 \, GeV$ )  which is two orders of magnitude higher than the earlier case.  The trajectory of the particle is shown in sub-figure 4(a) ( blue curve ), which clearly shows that in the presence of radiation reaction the particle passes through the focal point. Further the trajectory of the particle clearly exhibits the Doppler shift in the frequency of the wave seen by the particle. Since the parameter $\Delta$ is related to the Doppler shifted frequency as $\omega^{'} = \Delta \omega_{0}$, the rise and fall of $\Delta$ ( blue curve in Fig. 3(d) ) is reflected in the increase and decrease in oscillation frequency observed in the particle trajectory ( blue curve in Fig. 4(a) ).  We also note, that the dynamics of the charged particle in the presence of radiation reaction as governed by Landau-Lifshitz equation ( represented by the magenta curve ) shows an excellent match with that obtained using Hartemann-Luhmann equation, implying that the effect of Schott term in the radiation reaction dominated regime is negligible.  \\

For the second set of initial conditions, the particle dynamics is shown in Figs. (5) and (6). In Fig. (5), the sub-figures (a), (b), (c) and (d) respectively represent the longitudinal and transverse component of particle momentum, the energy and the parameter $\Delta$; and Fig. (6) represents the trajectory of the particle.
As before, we first discuss the particle dynamics in the absence of radiation reaction effects. The red curves in Fig. 5(a) and (b) respectively show that the average longitudinal momentum decreases 
( values correspond to y-axis on the left ) whereas the amplitude of the transverse momentum increases, as the particle approaches the focal point. The peak value of transverse momentum at the focal point matches 
$a_0$ ( $\mid \vec{p_{\perp}} \mid = a_0 = 1000$ ), in agreement with Eq. (\ref{eqn6}). As expected, in this case, the particle passes through the focal point, and the trend of average longitudinal momentum and amplitude of transverse momentum is reversed as it  moves away from the focal point. The particle gains average longitudinal momentum and loses transverse momentum in the defocused region. These observations are in agreement with  Eqs. (\ref{eqn8}) and (\ref{eqn6}). The corresponding energy gain is shown in Fig. 5(c) ( red curve with values corresponding to y-axis on the left ), which shows that for the second set of initial conditions, the average energy gain is around $\gamma \sim 700$ ($\sim 0.35 \, GeV$). Again we note that the average energy gain does not change throughout the motion ($\overline{d \gamma/d \tau} \approx 0$), which is in agreement with Eq. (\ref{energyeq_rr}). Finally Fig. 5(d), shows the evolution of the parameter $\Delta$ as a function of particle position 
$z$ ( red curve with values corresponding to y-axis on the left ). It monotonically increases upto the focal point and then monotonically decreases in the defocused region. The green curve on the top of the red curve is the plot of analytical expression of $\Delta$ as given by 
Eq. (\ref{Delta}), which clearly shows an excellent agreement with the numerical result. The red curve in Fig. (6), which represents the trajectory in the absence of radiation reaction effects ( values correspond to y-axis on the left ) shows, as expected, passage through the focal point.\\
  
With the second set of initial conditions, the particle dynamics in the presence of radiation reaction effects, is represented by the blue curve in Figs. (5) and (6) with values corresponding to y-axis on the right in sub-figures Fig. 5(a), 5(c), 5(d) and Fig. (6). It is found that the particle dynamics is qualitatively similar to that obtained using the first set of initial conditions. The energy gain, as with the first set of initial conditions, is found to be two orders of magnitude larger than that obtained without radiation reaction effects. Thus, as mentioned in the introduction,  irrespective of the choice of initial conditions, radiation reaction forces pushes the particle through the high intensity focal point, and the particle gains energy due to gain in forward longitudinal momentum. In this case also, the particle dynamics as governed by Landau-Lifshitz equation ( represented by magenta curve ) shows an excellent match with that obtained using Hartemann-Luhmann equation, again indicating that the effect of Schott term in the radiation reaction dominated regime is negligible.

\section{Summary and conclusions}

We have studied the dynamics of a charged particle in a focused light wave by taking account of radiation reaction effects. The dynamics have been studied using two well known equations {\it viz.} the Hartemann-Luhmann equation and the Landau-Lifshitz equation. We firstly show that both the equations give identical results and secondly, irrespective of the choice of initial conditions, in the presence of radiation reaction, the particle does not reflect from the focal region, thereby gaining a large amount of energy and forward momentum from the focused light wave. This result is in sharp contrast to the well known result by Kaw et. al.\cite{Kaw_Kulsrud}, derived in the absence of radiation reaction effects, where for certain initial conditions, the particle reflects from the high intensity region, thereby losing forward energy. The energy gain observed in the presence of radiation reaction, for our set of parameters, is found be two orders of magnitude greater than that obtained without radiation reaction effects; and also this result is independent of the choice of model equation. We have also found that these results are qualitatively the same when studied with a light pulse having a Gaussian envelope. Details of the study for a particle interacting with a light pulse having a Gaussian envelope including radiation reaction effects will be reported in a future publication.

\label{conclusion}
\nocite{*}  
\bibliography{focused}
\begin{figure}
  	\centering
 		{\includegraphics[width = 7in]{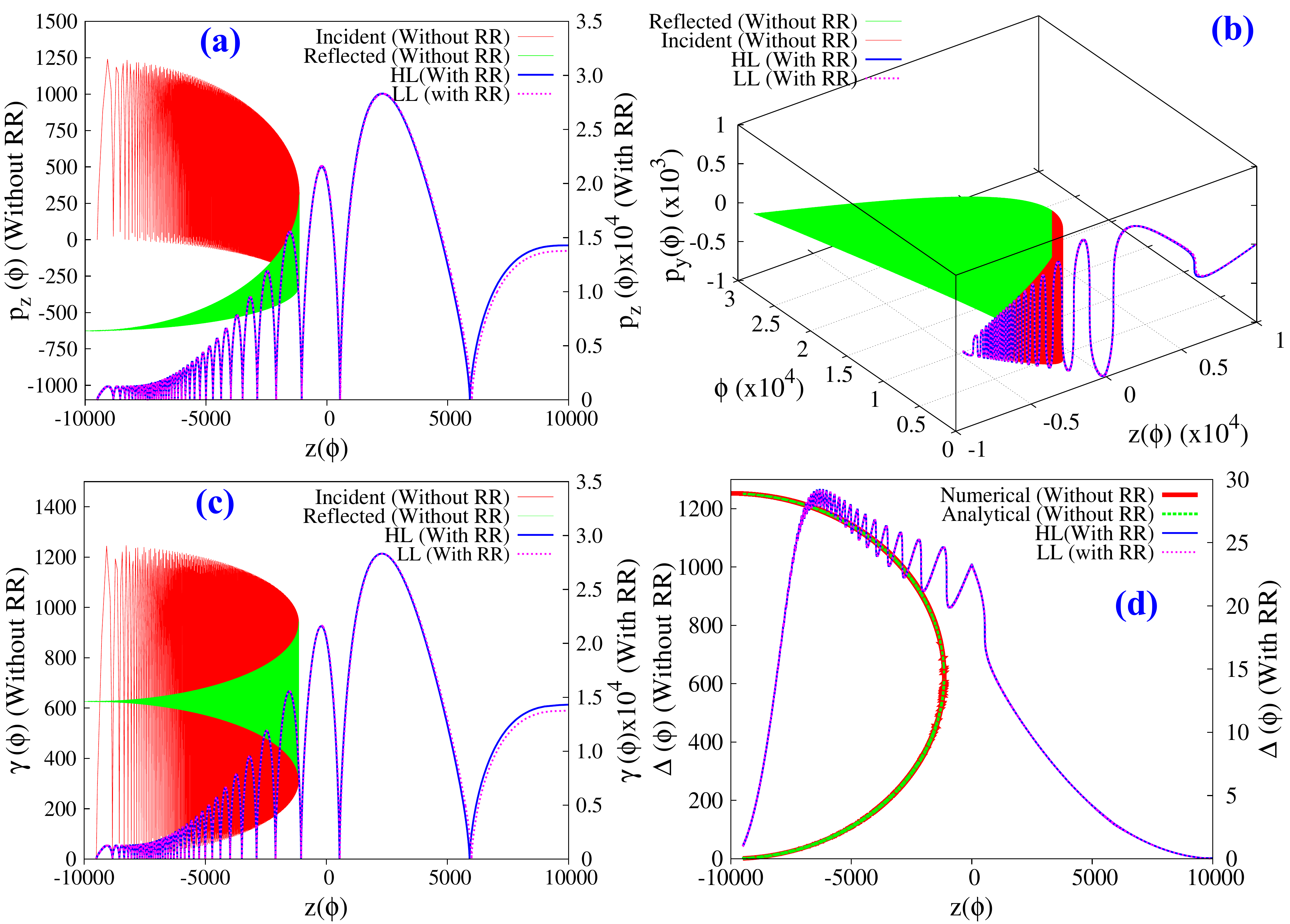}}
 	\caption{(a) - (d) respectively represent the evolution of longitudinal and transverse momentum, the energy and the parameter $\Delta$, for a charged particle interacting with a linearly polarized ($\delta = 0$), intense, focused electromagnetic wave, in the absence / presence of radiation reaction effects. The intensity at the focal point is chosen as  $a_0 = 1000$ ( $\sim 10^{24}\, W /cm^2$ ) and 
$\tau_0 \approx 1.8 \times 10^{-8}$. The initial conditions are 
$z_{0} = -9500$, $\vec{p} = 0$, $P_{\perp 0} = 0$ and $a(z_0) = 50$, which satisfy the inequality given by Eq. (\ref{con1}) ( $a_{z_{ref}} = 885.3 < a_0$; see text ). The red and green curves in Fig. 3(a) - (c) respectively represent the forward and reflected motion of the particle in the absence of radiation reaction effects ( values correspond to y-axis on the left ), and the corresponding value of the parameter $\Delta$ is represented in Fig. 3(d) where the red and green curve respectively represent the numerical values and the analytical expression for $\Delta$. The blue and magenta curves in the sub-figures (a) - (d) respectively represent the dynamics in the presence of radiation reaction effects ( values correspond to y-axis on the right ) as governed by Hartemann-Luhmann and Landau-Lifshitz equation of motion.}
\end{figure}
\begin{figure}
	\centering
	{\includegraphics[width = 5in]{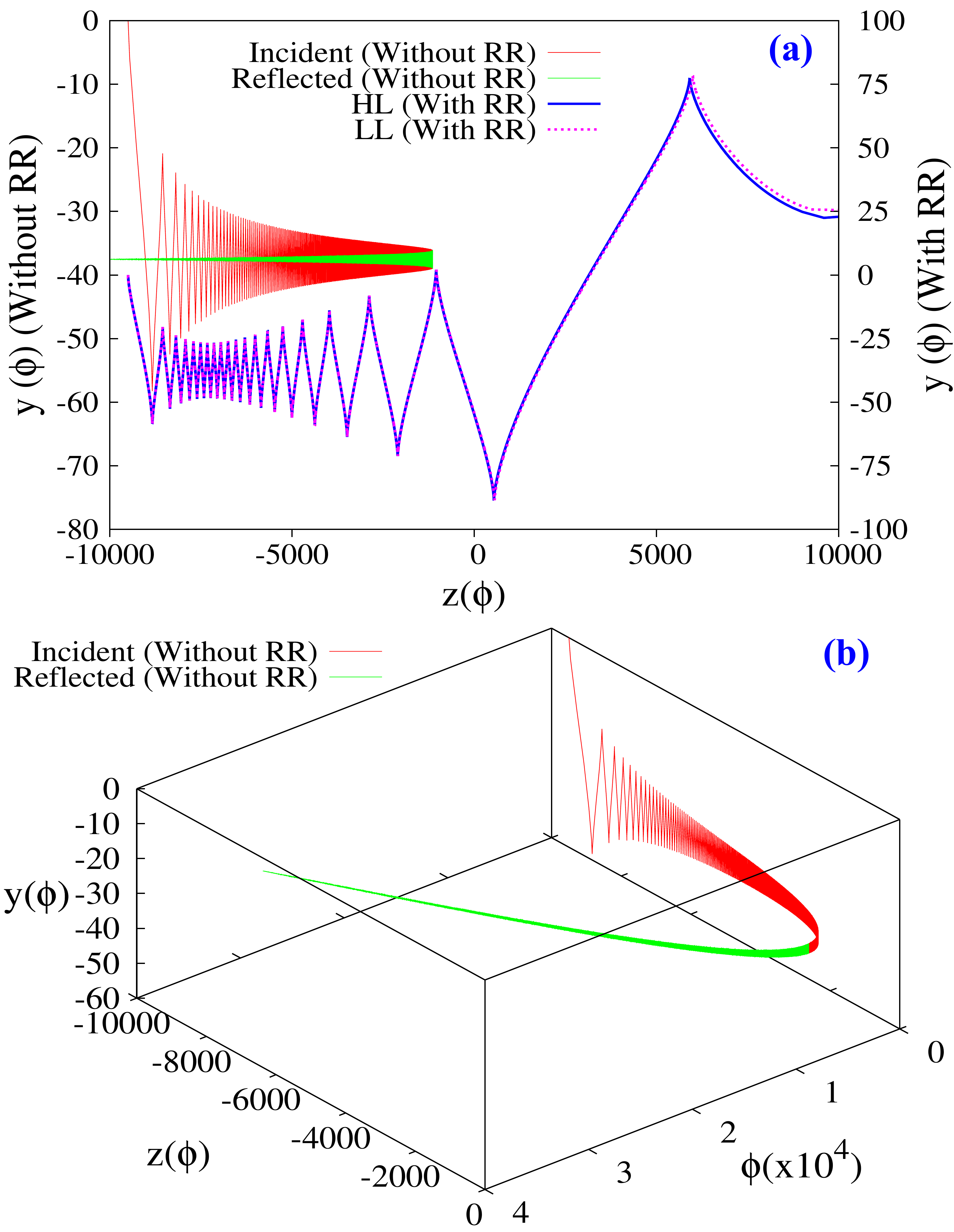}}
	\caption{ (a) -(b) represent the trajectory of the particle in the absence / presence of radiation reaction effects. The chosen initial conditions and the laser parameters are same as in Fig. (3). The red and green curves respectively represent the forward and reflected motion of the particle in the absence of radiation reaction  ( values correspond to y-axis on the left ). The blue and magenta curve in sub-figure 4(a) respectively represent the trajectory of the particle in the presence of radiation reaction effects ( values correspond to y-axis on the right )
as obtained by solving the Hartemann-Luhmann and the Landau-Lifshitz equation of motion}
\end{figure}
\begin{figure}
	\centering
	{\includegraphics[width = 7in]{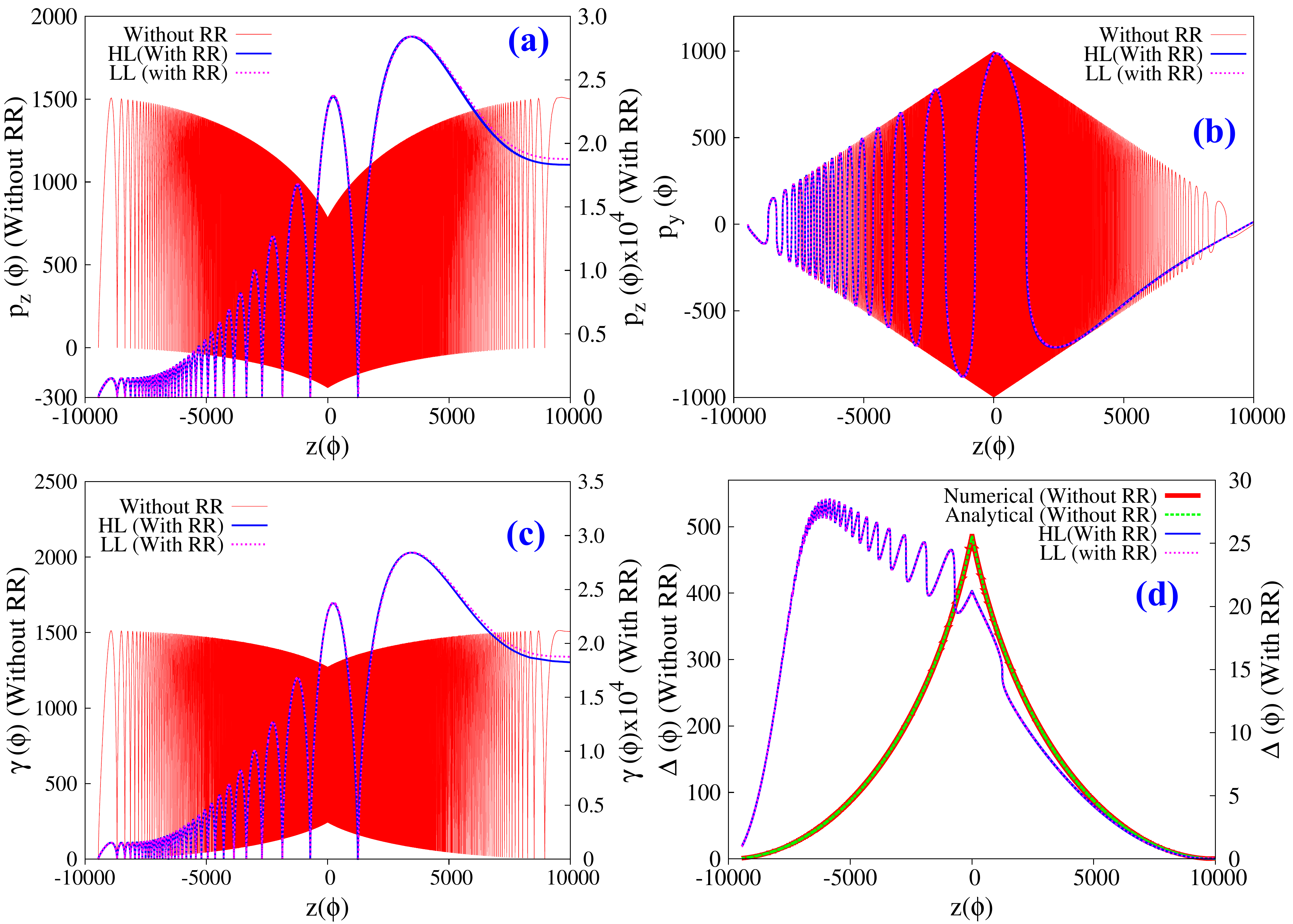}}
	\caption{(a) - (d) respectively represent the evolution of longitudinal and transverse momentum, the energy and the parameter $\Delta$, for a charged particle interacting with a linearly polarized ($\delta = 0$), intense, focused electromagnetic wave, in the absence / presence of radiation reaction effects. The intensity at the focal point is chosen as  $a_0 = 1000$ ( $\sim 10^{24}\, W /cm^2$ ) and 
$\tau_0 \approx 1.8 \times 10^{-8}$. The initial conditions are 
$z_{0} = -9450$, $\vec{p} = 0$, $P_{\perp 0} = 0$ and $a(z_0) = 55$, which violate the inequality given by 
Eq. (\ref{con1}) ( $a_{z_{ref}} = 1070.9 > a_0$; see text ). The red curve in Fig. 3(a) - (c) represent the motion of the particle in the absence of radiation reaction effects ( values correspond to y-axis on the left ), and the corresponding value of the parameter $\Delta$ is represented in Fig. 3(d) where the red and green curve respectively represent the numerical values and the analytical expression for $\Delta$. The blue and magenta curves in the sub-figures (a) - (d) respectively represent the dynamics in the presence of radiation reaction effects ( values correspond to y-axis on the right ) as governed by Hartemann-Luhmann and Landau-Lifshitz equation of motion.}
	\label{3_2}
\end{figure}
\begin{figure}
	\centering
	{\includegraphics[width = 7in]{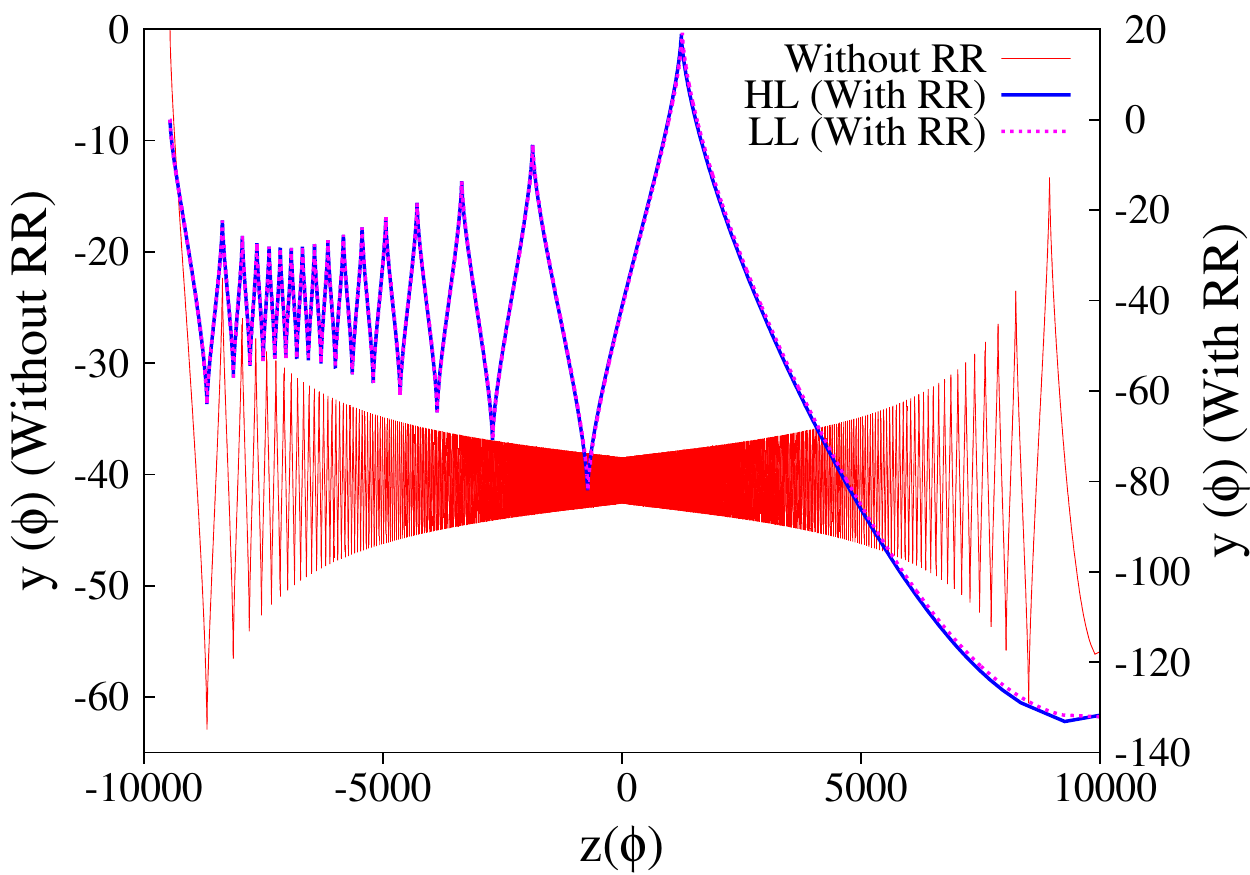}}
	\caption{ represent the trajectory of the particle in the absence / presence of radiation reaction effects. The chosen initial conditions and the laser parameters are same as in Fig. (5). The red curve represent the motion of the particle in the absence of radiation reaction  ( values correspond to y-axis on the left ). The blue and magenta curve respectively represent the trajectory of the particle in the presence of radiation reaction effects ( values correspond to y-axis on the right )
as obtained by solving the Hartemann-Luhmann and the Landau-Lifshitz equation of motion. }
	\label{3_3}
\end{figure}

\vspace{0.5cm}

\nocite{*}

\end{document}